\documentclass[a4paper,twoside,twocolumn,english,showpacs]{revtex4}

\usepackage{times}
\usepackage{amssymb}
\usepackage{graphicx}
\usepackage{amsmath}
\usepackage{color}

\newcommand{\note}[1]{}

\newcommand{\rC}{\text{\bf{C}}} 
 
\newcommand{\ecut}{\epsilon_{\rm cut}} 
\newcommand{\tecut}{\tilde{\epsilon}_{\rm cut}} 
\newcommand{\sumb}{\sum^-}

\newcommand{\PC}{\mathcal{P}_{}}

\begin{document}

\title{Numerical method for evolving the projected Gross-Pitaevskii equation }

\author{P. Blair Blakie}
\affiliation{  Jack Dodd Centre for Quantum Technology, Department of Physics, University of Otago, Dunedin,
New Zealand}

\date{\today}

\begin{abstract} 
In this paper we describe a  method  for evolving the projected Gross-Pitaevskii equation (PGPE) for a Bose gas in a harmonic oscillator potential. The central difficulty in solving this equation is the requirement that the classical field is restricted to a small set of prescribed modes that constitute the low energy classical region of the system.
We present a scheme, using a Hermite-polynomial based spectral representation, that precisely implements this mode restriction and allows an efficient and accurate solution of the PGPE. We show equilibrium and non-equilibrium results from the application of the PGPE to an anisotropic trapped three-dimensional Bose gas.
\end{abstract}

\maketitle
 \section{Introduction}
 
 Recently a variety of classical field methods have become popular in the description of ultra-cold Bose gases \cite{Steel1998a,Marshall1999a,Sinatra2001a,Goral2001a,Davis2001a,Davis2001d,Sinatra2002a,Davis2002a,Gardiner2002a,Gardiner2003a,Lobo2004a,DavisTemp,Norrie2004a,Polkovnikov2004a,Bradley,Blakie2005a,Davis2005a,Davis2006a,Simula2006a,Bezett2008a}. The appeal of these methods is that the dynamics of the modes are treated non-perturbatively so that non-equilibrium situations (e.g.~see \cite{Norrie2004a}) or strongly fluctuating equilibrium systems (e.g.~see \cite{Davis2006a}) can be accurately simulated. 
  
In Refs.  \cite{Davis2001a,Blakie2005a} we have developed a classical field theory, known as the Projected Gross Pitaevskii Equation (PGPE) formalism, to describe the finite temperature Bose gas. This approach has found good agreement with experiment in the critical region of the condensation transition \cite{Davis2006a}, and has seen numerous applications to regimes where traditional meanfield methods are inapplicable (e.g.~see \cite{Simula2006a,Bezett2008a,Simula2008a}).
A key component of our theory (and the primary distinction from other finite temperature classical field theories  \cite{Goral2001a}) that enables it to be applied to the quantitative description of experiments is the use of a projector, i.e. the explicit restriction of our description to the low energy modes of the system. 
For typical regimes of interest  of order a thousand modes of the system are sufficiently highly occupied to  be treated using a classical field approach \cite{Blakie2007a}. 


Over the past decade there has been extensive development of techniques for finding ground state solutions to the Gross-Pitaevskii equation and algorithms for evolving the condensate. The basic premise here is that the system is at zero temperature so that a single mode of the system is occupied and the underlying basis states used for the simulation are unimportant as long as they span  the condensate.  As such, in general these methods are not immediately applicable to the finite temperature case.

In Ref.~\cite{Blakie2005a} we have outlined a theoretical scheme for simulating a finite temperature trapped Bose gas. In this paper we present our algorithm for that case as well as the uniform system, and discuss the techniques we use to initialize and analyze PGPE simulations. As the typical usage of the PGPE formalism is far removed from the well-known zero temperature Gross-Pitaevskii equation (used to describe pure condensate dynamics), we use the results section to present a detailed example of application and analysis for the technique. This should be of benefit for others in the community wishing to adopt these techniques and also provides thermal quantities for comparison that arise from well-characterized simulations. An important result we present is a demonstration of irreversible behavior of the trapped PGPE, giving evidence for re-thermalization of this system.  

The outline of this paper is as follows. In the remainder of this section we briefly introduce the PGPE evolution equation. In Sec.~\ref{FormalAlgorithm} we set up a convenient set of units and outline our generic spectral approach to solving the PGPE equation in a finite basis. The details of implementing the algorithm are first presented for a uniform system in Sec.~\ref{SECnumunifo}. In this case the modes are plane-wave-like and the algorithm can be efficiently implemented using Fourier transforms. In Sec.~\ref{SEC:numharm} the main result of the paper is presented: An implementation for the experimentally relevant case of a harmonically trapped system. Here the natural modes to work in are the harmonic oscillator eigenstates, and we show that a finite number of such modes can be propagated accurately and efficiently using appropriate quadrature grids to evaluate the nonlinear matrix  elements.  
 Results for the finite temperature evolution of a harmonically trapped system are presented in Sec.~\ref{Results} before we conclude.

 \subsection{PGPE theory}
\begin{figure}
\includegraphics[%
  width=6.0cm,
  keepaspectratio]{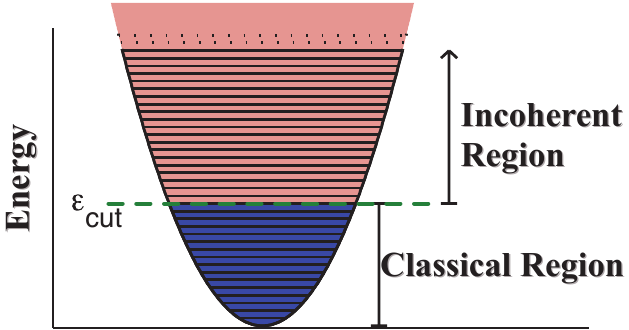}
\caption{\label{Fig:Classicalregion} Schematic diagram showing the classical
and incoherent regions of the single particle spectrum for a harmonically
trapped Bose gas. The energy $\epsilon_{{\rm cut}}$ is usually chosen so that the average
number of particles in the modes at the cutoff is   $n_{{\rm cut}}\sim1$.}
\end{figure}

 The  PGPE  is a time-dependent nonlinear Schr\"odinger equation of the form \cite{Davis2001a,Davis2002a,Blakie2005a}:
\begin{eqnarray}
i\hbar\frac{\partial\psi}{\partial t} & = &
H_{\rm{sp}}\psi 
 +\, \mathcal{P}\bigg\{ NU_0|\psi|^{2}\psi\bigg\},\label{eq:PGPE}\end{eqnarray}
where
\begin{eqnarray}
H_{\rm{sp}}&=& H_0+\delta V(\mathbf{x},t),\\
H_0&=& 
-\frac{\hbar^{2}}{2m}\nabla^{2}+V_{{\rm trap}}(\mathbf{x}).
\end{eqnarray} 
$H_{\rm{sp}}$ is the single particle Hamiltonian, which includes the dominant $H_0$ part and we allow for a (small) perturbation part $\delta V$,  $\psi=\psi(\mathbf{x},t)$ is the classical matter wave field (taken to be normalized to unity), $N$ is the total number of atoms described by the PGPE, $V_{{\rm trap}}(\mathbf{x})$
is the external trapping potential, and $U_{0}=4\pi a\hbar^{2}/m$,
with $a$ the s-wave scattering length.
The numerical implementation of the PGPE formalism poses a rather interesting challenge: Formally only the low energy modes of the system are classical (i.e. the classical region shown schematically in Fig.~\ref{Fig:Classicalregion} \cite{Blakie2007a}) and should be the only modes retained in the numerical description, a  restriction expressed formally by the projector 
 \begin{equation}
\mathcal{P}\{ F(\mathbf{x})\}\equiv\sum_{n\in\rC}\phi_{n}(\mathbf{x)}\int d^{3}\mathbf{x}'\phi_{n}^{*}(\mathbf{x'})F(\mathbf{x'}),\label{eq:projector}\end{equation}
where $\phi_{n}(\mathbf{x})$ are eigenstates of  $H_{0}$
and the summation is restricted to modes in the classical region.
The action of $\mathcal{P}$ in Eq.~(\ref{eq:projector}) is thus to project
the arbitrary function $F(\mathbf{x})$ into the classical region ($\rC$), which we take to be defined by the single particle eigenstates up to some specified (single particle) energy $\epsilon_{\rm cut}$.
For a full description of the system dynamics we will require a means of simulating the incoherent region (i.e. the modes complementary to the classical region -- see Fig.~\ref{Fig:Classicalregion}) and a means to couple the two regions. 
A theoretical formalism for doing this has been presented in Ref. \cite{Gardiner2003a}, but has yet to be fully implemented numerically. However, the purpose of this paper is to present our approach for efficiently and accurately simulating Eq.~(\ref{eq:PGPE}) for the experimentally relevant case of a harmonic oscillator external potential.

\section{Formal algorithm}\label{FormalAlgorithm}
\subsection{Numerical requirements}
The modes of the system are of central importance in the implementation of the PGPE and care must be taken in numerical implementations to ensure the modes are
faithfully represented. Any useful
simulation technique must satisfy the following requirements.
\begin{itemize}
\item[(i)]The space spanned by the modes of the simulation
should match that of the classical region of the physical system
being simulated as closely as possible. That is, the
modes should be the single-particle modes of the system (i.e. eigenstates of $H_0$) up
to the prescribed energy cutoff  $\ecut$.
\item[(ii)] The assumption of high occupancy in all modes necessitates
that the numerical scheme must propagate all
modes accurately.
\end{itemize}
Most commonly used methods for propagating
Schr\"odinger-type equations do not satisfy these requirements;
in particular, many methods do not propagate all
modes of the numerical basis faithfully. This leads to negligible
errors if the highest modes are unoccupied, as is the
case for the $T=0$  Gross-Pitaevskii equation. However, it is clear that methods
based on such assumptions will not be appropriate for simulating
fields where the dynamics of all the modes are important.

\subsection{Computational units}
For convenience we present the discussion of our numerical methods in computational units, indicated by tildes. We do this by introducing appropriate units of distance $x_0$ and time $t_0$ in each of the following sections \footnote{These choices immediately imply computational units for  energy $E_0=\hbar/t_0$ and wavevector $k_0=1/x_0$ and hence momentum $p_0=\hbar/x_0$.}. So for example, our dimensionless distance variable is defined as $\tilde x = x/x_0$, dimensionless time is $\tilde t=t/t_0$, and classical field $\tilde\psi = \psi x_0^{3/2}$. The coefficient of the nonlinear term in the Gross-Pitaevskii equation is given by the product $NU_0$. In dimensionless units we define this as the nonlinearity constant $C\equiv NU_0t_0/\hbar x_0^3$.

Adopting computational units, the projected Gross-Pitaevskii equation takes the form
\begin{eqnarray}
i\frac{\partial\tilde\psi}{\partial \tilde t} & = &\PC\left\{ \tilde H_{\rm sp}\tilde\psi+ C |\tilde\psi|^{2}\tilde\psi\right\}.\label{eq:numGPE1}\end{eqnarray}

\subsection{Implementing the projector}
The projector can be implicitly implemented by restricting the classical field to the modes of interest, i.e.
\begin{equation}
\tilde{\psi}(\tilde{\mathbf{x}},\tilde t)=\sum_{n\in\rC}c_{n}(\tilde t)\,\tilde\phi_{n}(\tilde{\mathbf{x}}),\label{eq:psibasis}\end{equation}
 where $\{\tilde\phi_{n}(\tilde{\mathbf{x}})\}$ are the eigenstates of  $\tilde H_0$ with respective eigenvalue $\tilde\epsilon_{n}$. The projection
is effected by limiting the summation indices in Eq.~(\ref{eq:psibasis}) to the set of values
\begin{equation}
\rC=\{n:\tilde\epsilon_{n}\leq \tecut\},\label{eq:Cset}\end{equation}  
i.e. the field $\tilde\psi$ only contains the modes of interest. 

\subsubsection{Mode evolution}
Having restricted the modes for the purposes of the projector, we can adopt these modes as our spectral basis and  represent their evolution exactly using a Galerkin approach  (i.e. projecting   Eq.~(\ref{eq:numGPE1}) on to our spectral basis). This leads to an evolution equation for the amplitudes
\begin{eqnarray}
\frac{\partial c_{n}}{\partial\tilde t} & = & -i\left[\tilde\epsilon_{n}c_n+F_n+CG_{n}\right],\label{eq:GPEshobasis}\end{eqnarray}
 where\begin{eqnarray}
 F_n &\equiv&\int d^{3}\tilde{\mathbf{x}}\:\tilde\phi_{n}^{*}(\tilde{\mathbf{x}})\widetilde{\delta V}(\tilde{\mathbf{x}},\tilde t)\tilde\psi(\tilde{\mathbf{x}},\tilde{t}),\label{eq:Fpert}\\
G_{n}&\equiv&\int d^{3}\tilde{\mathbf{x}}\:\tilde\phi_{n}^{*}(\tilde{\mathbf{x}})|\tilde\psi(\tilde{\mathbf{x}},\tilde t)|^{2}\tilde\psi(\tilde{\mathbf{x}},\tilde{t}),\label{eq:GNL}\end{eqnarray}
are the matrix elements of the perturbation potential and nonlinear term respectively. Once all the matrix elements on the right hand side of Eq.~(\ref{eq:GPEshobasis}) are evaluated, the evolution of the system can be calculated using  numerical algorithms for  systems of ordinary differential equations, e.g. the Runge-Kutta algorithm (e.g. see \cite{Press92a}). Here we do not concern ourselves with the particular choice of propagation algorithm, but instead focus on evaluating the matrix elements. We do mention in passing that by moving to an \emph{interaction picture}, defined by the transformation $\bar{c}_n(\tilde{t})=\exp(i\tilde{\epsilon}_n\tilde{t})c_n(\tilde{t})$, the explicit dependence on $\tilde{\epsilon}_n$ can be removed from Eq.~(\ref{eq:GPEshobasis}) (also see \cite{Bao2005a}).


\section{Implementation for a uniform system\label{SECnumunifo}}
Here we consider the numerical description of a Bose gas in a cuboid volume with linear dimensions $\{L_x,L_y,L_z\}$ and subject to periodic boundary conditions. To simplify our discussion we consider the case where $L_{x}=L_{y}=L_{z}$ so that the 1D basis states (see below) are identical in each direction.
For a Bose gas in a uniform system with periodic boundary conditions
the basis Hamiltonian takes the form
\begin{equation}
\tilde{H}_0=-\frac{{\tilde\nabla}^{2}}{(2\pi)^2},\label{EQ:H0UniformNumerical}
\end{equation}
with the boundary conditions,
\begin{equation}
\tilde{\psi}(\tilde{x}+1,\tilde{y},\tilde{z})=\tilde{\psi}(\tilde{x},\tilde{y}+1,\tilde{z})=\tilde{\psi}(\tilde{x},\tilde{y},\tilde{z}+1)=\tilde{\psi}(\tilde{x},\tilde{y},\tilde{z}),\label{EQ:unifBDs}
\end{equation}
where we have taken $x_{0}=L_{x}$ as the unit of length and $t_{0}=mL_{x}^{2}/\pi h$
as the unit of time.
\subsection{Separating into 1D basis states}
The basis states (eigenstates of $H_0$) are separable into 1D eigenstates, i.e.
\begin{eqnarray}
\tilde{\phi}_n(\tilde{\mathbf{x}})&\leftrightarrow&\tilde\varphi_{\alpha}(\tilde{x})\tilde\varphi_{\beta}(\tilde{y})\tilde\varphi_{\gamma}(\tilde{z}),\label{EQ:PW1Dstates}\\
\tilde{\epsilon}_n&\leftrightarrow&\tilde{\varepsilon}_{\alpha}+\tilde{\varepsilon}_{\beta}+\tilde{\varepsilon}_{\gamma},\label{EQ:PW1Denergies}\\
c_n&\leftrightarrow&c_{\alpha\beta\gamma}, 
\end{eqnarray} 
where \begin{equation} 
\tilde{\varphi}_{\alpha}(\tilde{x})=e^{i\tilde{k}_{\alpha}\cdot\tilde{x}},\label{EQ:pwstates1D}\end{equation}
are the 1D eigenstates of the kinetic energy operator with the wavevectors
$\tilde{k}_{\alpha}$ chosen as harmonics of the perodicity interval, i.e.
$\tilde{k}_{\alpha}=2\pi \alpha,$ with $\alpha$ an integer, and have
respective eigenvalues $\tilde{\varepsilon}_{\alpha}=\alpha^{2}.$ 
For clarity we use Greek subscripts to label the 1D eigenstates, and note that the classically simulated region limits the values these can take to the set \begin{equation}
\rC=\{\alpha,\beta,\gamma:\tilde\varepsilon_{\alpha}+\tilde\varepsilon_{\beta}+\tilde\varepsilon_{\gamma}\le\tecut\},
\end{equation}
i.e. a sphere of radius $\sqrt{\tecut}$ in $\alpha\beta\gamma$-space.
For later convenience
we define $\alpha_{\max}$ as the maximum value of $\alpha$  that occurs in $\rC$, i.e. the highest order  basis state in each direction. For the planewave case we have $\alpha_{\max}\simeq\sqrt{\tecut}$. This means that within the classical region there exists $M=2\alpha_{\max}+1$ distinct 1D eigenstates (i.e. $\tilde{\varphi}_{\alpha}$) in each direction (since $\alpha\in[-\alpha_{\max},\alpha_{\max}]$), and  $M_T\approx \frac{\pi}{6}M^3\,$ 3D basis states ($\tilde{\phi}_n$) in the classical region. 
 In what follows, we will adopt the notation $\bar{\sum}$ to indicate summations restricted to the classical region, i.e. all triplets of Greek indices in $\rC$.

\subsection{Evaluating the matrix elements}\label{oscimatelem}
The nonlinear matrix element (\ref{eq:GNL}) takes the form  \begin{equation}
G_{\alpha\beta\gamma}\equiv\int d^{3}\tilde{\mathbf{x}}\,\tilde{\varphi}_{\alpha}^{*}(\tilde{x})\tilde{\varphi}_{\beta}^{*}(\tilde{y})\tilde{\varphi}_{\gamma}^{*}(\tilde{z})\left|\tilde{\psi}\left(\tilde{\mathbf{x}},\tilde{t}\right)\right|^{2}\tilde{\psi}\left(\tilde{\mathbf{x}},\tilde{t}\right).\label{EQ:GNLPW}\end{equation}
Substituting the basis expansion for the field, given by Eqs. (\ref{eq:psibasis}) and (\ref{EQ:PW1Dstates}) into Eq.~(\ref{EQ:GNLPW}) gives a series of matrix elements that can be independently integrated over the coordinate directions. Each of these 1D integrals is of the form
\begin{equation}
I_{\alpha\delta\nu\sigma}\equiv\int d\tilde{x}\,\tilde{\varphi}_{\alpha}^{*}(\tilde{x})\tilde{\varphi}_{\delta}^{*}(\tilde{x})\tilde{\varphi}_{\nu}(\tilde{x})\tilde{\varphi}_{\sigma}(\tilde{x}),\label{Imnrq}
\end{equation}
which the indices in general can take any of the $M$ values in the range  $-\alpha_{\max},-(\alpha_{\max}-1),\ldots,\alpha_{\max}$.
These integrals are in some sense trivial for the planewave case, i.e. $I_{\alpha\delta\nu\sigma}=\delta_{\alpha+\delta,\nu+\sigma}$, however knowledge of these matrix elements is not sufficient to implement an efficient algorithm, as we discuss below.

Expanding the field $\tilde{\psi}$ in Eq.~(\ref{EQ:GNLPW}) in terms of the basis states, the nonlinear matrix elements can be written as 
\begin{equation}
G_{\alpha\beta\gamma}=
\sumb_{\delta\zeta\eta}\sumb_{\nu\xi\rho}\sumb_{ \sigma\tau\upsilon}
c^*_{\delta \zeta \eta}c_{\nu\xi\rho}c_{\sigma\tau\upsilon}I_{\alpha\delta\nu\sigma}I_{\beta\zeta\xi\tau}I_{\gamma\eta\rho\upsilon}.\label{EQ:GNLsum}
\end{equation}
Explicitly carrying out the summations in Eq.~(\ref{EQ:GNLsum}) for all the matrix elements of $G_{\alpha\beta\gamma}$ requires $O(M^{12})$ operations, and would be prohibitively slow for any practical calculation.  We now show how a quadrature approach can be used to evaluate these integrals much more efficiently, requiring only $O(M^4)$ operations. The essence of this approach is to transform the field to a spatial representation where the nonlinear term is local. By choosing the spatial grid as an appropriate quadrature grid, the matrix elements will still be evaluated exactly.
The basic idea of this approach (efficient evaluation of nonlinear matrix elements on a spatial grid) is widely used (e.g. see \cite{Schneider1999a,Feder2000a,Roi2000a,Collecutt2001a,Dion2003a,Bao2005a}), although it is usually implemented in a manner that is not exact. This is typically an acceptable approximation if the highest energy modes of the system are unoccupied, a luxury not available in the PGPE.

In each spatial dimension, the quadrature grid of interest (for the uniform case) consists of $N_{\rm Q}$-points  given by \begin{equation}
\tilde{x}_{j}=j\,\Delta\tilde{x},\qquad1\le j\le N_{\rm Q},\label{EQ:PWxgrid}\end{equation}
with spacing $\Delta\tilde{x}=1/N_{\rm Q}$, which spans the spatial region $(0,1]$. The quadrature expression for an integral of an arbitrary  function $f$ is
\begin{equation}
\int_a^bd\tilde{x}\,w(\tilde x)\,f(\tilde x)\approx\sum_{j=1}^{N_{\rm Q}} w_j f(\tilde x_j),
\end{equation} 
where $w(\tilde x)$ is the weight function, and $w_j$ are the quadrature weights.
Formally such a quadrature can be implemented for our  planewave case if we take $a=0$, $b=1$, $w(\tilde{x})=1$ and $w_j=\Delta\tilde{x}$.
 
 The requirement that our quadrature will exactly calculate the nonlinear matrix elements is equivalent to the requirement that the 1D matrix elements (\ref{Imnrq}) are all evaluated exactly, i.e.
\begin{eqnarray}
I_{\alpha\beta\gamma\delta}&=&\sum_{j=1}^{N_{\rm Q}}\Delta\tilde{x}\,\tilde{\varphi}_{\alpha}^{*}(\tilde{x}_{j})\tilde{\varphi}_{\beta}^{*}(\tilde{x}_{j})\tilde{\varphi}_{\gamma}(\tilde{x}_{j})\tilde{\varphi}_{\delta}(\tilde{x}_{j}),\\
&=&\delta_{\alpha+\beta,\gamma+\delta},\end{eqnarray}
which holds for the quadrature described above if we take $N_{\rm Q}\ge 2M$. For what follows we take   $N_{\rm Q}=2M$.
This quadrature can be applied to the 3D case, so that the nonlinear matrix elements are computed as
\begin{eqnarray}
G_{\alpha\beta\gamma}&=&\sum_{ ijk}(\Delta\tilde{x})^{3}\,\tilde{\varphi}_{\alpha}^{*}(\tilde{x}_i)\tilde{\varphi}_{\beta}^{*}(\tilde{y}_j)\tilde{\varphi}_{\gamma}^{*}(\tilde{z}_k)\nonumber\\
&&\times|\tilde{\psi}(\tilde{\mathbf{x}}_{ijk},\tilde{t})|^{2}\tilde{\psi}(\tilde{\mathbf{x}}_{ijk},\tilde{t}),\end{eqnarray}
 where $\tilde{\mathbf{x}}_{ijk}=(\tilde{x}_{i},\tilde{y}_{j},\tilde{z}_{k})$, the $y$ and $z$ grids are taken to be identical to the $x$ grid, and the summation over $\{i,j,k\}$ hereafter will be each taken to be over the range of values $1$ to $N_{\rm Q}=2M$. 
 
 Unless the perturbation potential can be expressed as a superposition of the  planewave states (\ref{EQ:pwstates1D}) of maximum wavevector $\tilde{k}_{\max}=4\pi \alpha_{\max}$ in each direction, the quadrature given above will not be exact for evaluating Eq.~(\ref{eq:Fpert}).  However, since the perturbation will normally be small this approximate treatment should be satisfactory, i.e. we will take
 \begin{eqnarray}
 F_{\alpha\beta\gamma}&\approx&\sum_{ ijk}{}(\Delta\tilde{x})^{3}\,\tilde{\varphi}_{\alpha}^{*}(\tilde{x}_i)\tilde{\varphi}_{\beta}^{*}(\tilde{y}_j)\tilde{\varphi}_{\gamma}^{*}(\tilde{z}_k)\nonumber\\
 &&\times\widetilde{\delta V}(\tilde{\mathbf{x}}_{ijk},\tilde{t}) \tilde{\psi}(\tilde{\mathbf{x}}_{ijk},\tilde{t}).
 \end{eqnarray}

\subsection{Overview of numerical procedure}\label{SEC:PWnummeth}
Here we briefly overview how the quadrature described above can be efficiently implemented numerically. Starting from the basis set representation of the field (i.e. $\{c_{\alpha\beta\gamma}\}$) at an instant of time $\tilde t$, the steps for calculating the matrix elements are as follows:
\begin{enumerate}
\item The field is transformed to a spatial representation according to 
\begin{equation}
\tilde{\psi}(\tilde{\mathbf{x}}_{ijk},\tilde t)=\sumb_{\alpha\beta\gamma}U_{i\alpha}U_{j\beta}U_{k\gamma}\,c_{\alpha\beta\gamma}(\tilde t),
\end{equation}
where the transformation matrices are defined as the 1D basis states evaluated on the quadrature grid, i.e.
\begin{equation}
U_{i\alpha}=\tilde\varphi_{\alpha}(\tilde{x}_i).
\end{equation}
\item The integrands of the matrix elements (\ref{eq:Fpert}) and (\ref{eq:GNL}) are then constructed, i.e.
\begin{eqnarray}
f(\tilde{\mathbf{x}}_{ijk})&\equiv&\widetilde{\delta V}(\tilde{\mathbf{x}}_{ijk},\tilde{t})\tilde{\psi}(\tilde{\mathbf{x}}_{ijk},\tilde{t}),\\
g(\tilde{\mathbf{x}}_{ijk})&\equiv&|\tilde{\psi}(\tilde{\mathbf{x}}_{ijk},\tilde{t})|^{2}\tilde{\psi}(\tilde{\mathbf{x}}_{ijk},\tilde{t}).
\end{eqnarray}

\item Inverse transforming these integrand functions yields the desired matrix elements:
\begin{eqnarray}
F_{\alpha\beta\gamma} &=& (\Delta \tilde x)^3\sum_{ijk}U_{i\alpha}^*U_{j\beta}^*U_{k\gamma}^*f(\tilde{\mathbf{x}}_{ijk}),\\
G_{\alpha\beta\gamma} &=& (\Delta \tilde x)^3\sum_{ijk}U_{i\alpha}^*U_{j\beta}^*U_{k\gamma}^*g(\tilde{\mathbf{x}}_{ijk}).
\end{eqnarray}
\end{enumerate}
The slowest step in this procedure is carrying out the basis transformation (steps 1 and 3). The computational cost of this step is $O(M^4)$ when carried out as a series of matrix multiplications. However, for the planewave case this transformation is equivalent to a fast Fourier transformation, which has a computational cost of $O\left(M^3\log(M)\right)$. In the last step the transforms to obtain $F$ and $G$ can be combined into a single transform, i.e. in practice we transform  $f(\tilde{\mathbf{x}}_{ijk})+Cg(\tilde{\mathbf{x}}_{ijk})$, which yields $F_n+CG_n$ (see  Eq.~(\ref{eq:GPEshobasis})).

\section{Implementation for harmonically trapped system\label{SEC:numharm}}
We now consider the main subject of this paper: The implementation of the PGPE for the harmonically trapped system, i.e. where
\begin{equation}
V_{\rm trap}(\mathbf{x})=\frac{1}{2}m\left(\omega_x^2x^2+\omega_y^2y^2+\omega_z^2z^2\right).
\end{equation}
In computational units the basis Hamiltonian, $\tilde{H}_0$, takes the form 
\begin{eqnarray}
\tilde{H}_0& = & -\frac{1}{2}\tilde\nabla^{2}+\frac{1}{2}(\lambda_{x}^{2}\tilde x^{2}+\lambda_{y}^{2}\tilde y^{2}+\tilde z^{2}), \label{eq:H0harm}\end{eqnarray}
where $\lambda_x=\omega_x/\omega_z$, $\lambda_y=\omega_y/\omega_z$, and we have used the harmonic oscillator frequency associated with the $z$-direction to define units of length $x_0=\sqrt{\hbar/m\omega_z}$ and time $t_0=\omega_z^{-1}$.
To simplify our discussion of the numerical method, we will
take the harmonic trapping potential to be isotropic, i.e. $\lambda_{x}=\lambda_{y}=1$.
This allows us to avoid using cumbersome notation to account for different
spectral bases in each direction.
 
 The same decomposition used for the planewave case (\ref{EQ:PW1Dstates}) can be applied in the harmonically trapped system if we take $\{\tilde\varphi_{\alpha}(\tilde x)\}$ to be eigenstates of the 1D harmonic
oscillator Hamiltonian, i.e.\begin{equation}
\left[-\frac{1}{2}\frac{d^{2}}{d\tilde x^{2}}+\frac{1}{2}\tilde x^{2}\right]\tilde\varphi_{\alpha}(\tilde x)=\tilde\varepsilon_{\alpha}\tilde\varphi_{\alpha}(\tilde x),\label{eq:sho1D}\end{equation}
 with eigenvalue $\tilde\varepsilon_{\alpha}=(\alpha+\frac{1}{2})$, where we have taken $\alpha$ to be a non-negative integer. Such spectral representations have been considered previously for the zero-temperature (non-projected) Gross-Pitaevskii equation in Refs. \cite{Dion2003a,Bao2005a}.

We define $\alpha_{\max}$ as the maximum value of $\alpha$ that occurs in $\rC$, i.e. for the harmonic case $\alpha_{\max}\simeq\tecut-\frac{1}{2}$. This means that within the classical region there exists $M=\alpha_{\max}+1$ distinct 1D eigenstates (i.e. $\tilde{\varphi}_{\alpha}$) in each direction (since $\alpha$ takes the values $0,1,\ldots, \alpha_{\max}$), and  $M_T\approx \frac{1}{6}M^3$ 3D basis states ($\tilde{\phi}_n$) in the classical region.

\subsection{Oscillator state properties}
We briefly review the properties of the harmonic oscillator states,
which we utilize as our spectral basis.

The eigenstates of the 1D single particle Hamiltonian (\ref{eq:sho1D}) are 
\begin{equation}
\tilde{\varphi}_{\alpha}(\tilde{x})=h_{\alpha}H_{\alpha}(\tilde{x})e^{-\tilde{x}^{2}/2},
\end{equation}
where $h_{\alpha}=[{2^{\alpha}\alpha!\sqrt{\pi}}]^{-1/2}$ is the normalization constant, and   $H_{\alpha}(\tilde x)$ is a Hermite polynomial of degree $\alpha$, defined by the recurrence relation
\begin{equation}
H_{\alpha+1}(\tilde{x})=2\tilde{x}H_{\alpha}(\tilde{x})-2\alpha H_{\alpha-1}(\tilde{x}),\quad \alpha=1,2,\ldots
\end{equation}
 with $H_{0}(\tilde{x})=1,$ and $H_{1}(\tilde{x})=2\tilde{x}$. 
 The harmonic oscillator states are eigenstates of the Fourier transform operator with eigenvalue $(-i)^{\alpha}$, i.e. 
 \begin{equation}
 \tilde{\varphi}_{\alpha}{(\tilde{p})}=(-i)^{-\alpha}\frac{1}{\sqrt{2\pi}}\int d\tilde{x}\,e^{-i\tilde{p}\tilde{x}}\tilde{\varphi}_{\alpha}(\tilde{x}).
 \end{equation}
 Thus knowledge of the basis amplitudes $c_{\alpha\beta\gamma}$ allows us to efficiently and precisely construct the momentum representation of the classical field, i.e.
 \begin{equation}
 \tilde{\Phi}(\tilde{\mathbf{p}})=\bar{\sum_{\alpha\beta\gamma}}(-i)^{\alpha+\beta+\gamma}c_{\alpha\beta\gamma}\,\tilde{\varphi}_{\alpha}(\tilde{p}_x)\tilde{\varphi}_{\beta}(\tilde{p}_y)\tilde{\varphi}_{\gamma}(\tilde{p}_z).\label{EQmtmfield}
 \end{equation}

\subsubsection{Step operators}\label{SEC:stepops}
It is useful to consider the so-called \emph{step operators} of quantum mechanics, defined as 
\begin{eqnarray}
\hat a_{x}^{+} & = & \frac{1}{\sqrt{2}}\left(-\frac{\partial}{\partial \tilde x}+\tilde x\right),\\
\hat a_{x}^{-} & = & \frac{1}{\sqrt{2}}\left(\frac{\partial}{\partial \tilde x}+\tilde x\right),\end{eqnarray}
which are mutually adjoint and have the commutation relation 
\begin{equation}
[\hat a_{x}^{-},\hat a_{x}^{+}]=1.\label{eq:acomm}
\end{equation}
In a similar manner we can define step operators for the $\tilde y$ and $\tilde z$ directions, although for the sake of brevity we avoid doing this here.

Applying the step operators to the 1D eigenstates yields
\begin{eqnarray}
\hat a_{x}^{+}\tilde{\varphi}_{\alpha}(\tilde x) &=& \sqrt{\alpha+1}\, \tilde{\varphi}_{\alpha+1}(\tilde x),\\
\hat a_{x}^{-}\tilde{\varphi}_{\alpha}(\tilde x) &=& \sqrt{\alpha}\, \tilde{\varphi}_{\alpha-1}(\tilde x),
\end{eqnarray}
so that the matrix representation of these operators in the spectral basis is\begin{eqnarray}
\left(\hat a_{x}^{+}\right)_{\alpha\beta} & \equiv & \int dx\,\tilde{\varphi}^*_{\alpha}(\tilde x) \,\hat a_{x}^{+}\,\tilde{\varphi}_{\beta}(\tilde x), \\
&=& \sqrt{\beta+1}\,\delta_{\alpha,\beta+1},\\
\left(\hat a_{x}^{-}\right)_{\alpha\beta} & = & \sqrt{\beta}\,\delta_{\alpha,\beta-1}.\end{eqnarray}
 Most importantly this allows us to represent the operators $\tilde x$ and $\partial_x\equiv\frac{\partial}{\partial \tilde x}$
in the spectral basis (exactly) as\begin{eqnarray}
(\hat{  x})_{\alpha\beta} & =&\frac{1}{\sqrt{2}}(\hat a_{x}^{+}+\hat a_{x}^{-})_{\alpha\beta}, \\ 
&=&\sqrt{\frac{\beta+1}{2}}\delta_{\alpha,\beta+1}+\sqrt{\frac{\beta}{2}}\delta_{\alpha,\beta-1},\\
\left( \widehat{\partial_x}\right)_{\alpha\beta} & =&\frac{1}{\sqrt{2}}(\hat a_{x}^{-}-\hat a_{x}^{+})_{\alpha\beta},\\ 
& =&\sqrt{\frac{\beta}{2}}\delta_{\alpha,\beta-1}-\sqrt{\frac{\beta+1}{2}}\delta_{\alpha,\beta+1}.\end{eqnarray} 
So, for example, consider the position expectation of the field, i.e. $\langle \tilde{x}(\tilde t)\rangle= \int d^3\tilde{\mathbf{x}}\,\tilde x |\tilde{\psi}(\tilde{\mathbf{x}},\tilde t)|^2$. Note that this expectation is a quantum mechanical average at time $\tilde{t}$, rather than a thermal (ensemble/time) average.
This quantity can be calculated in the spectral basis as
\begin{equation}
\langle \tilde x(\tilde t)\rangle =  \sum^{\sim}_{\alpha\beta\gamma\delta}c^*_{\delta\beta\gamma}(\tilde t)\,(\hat x)_{\delta\alpha}\,c_{\alpha\beta\gamma}(\tilde t),
\end{equation}
where $\tilde{\sum}_{\alpha\beta\gamma\delta}$ indicates a restricted summation over the variables $\{\alpha,\beta,\gamma,\delta\}$ such that both $\{\alpha,\beta,\gamma\}$ and $\{\delta,\beta,\gamma\}$ lie in $\rC$.
While this appears to be of computational cost $O(M^4)$, in fact the sparseness of the underlying step operators means that this operation is $O(M^3)$. 

Some care needs to be taken when applying the step operators. Formally, these operators have infinite size matrix representations, whereas our classical field application is intrinsically finite by virtue of the energy cutoff. An implication of this is that, e.g. the action of the $\hat a^+_x$ operator will take the highest  $x$-modes in the classical region onto modes above the cutoff, which would then be lost from our description. This is a problem when we wish to consider the product of operators and can be avoided by using the commutation relation (\ref{eq:acomm}) to write operators of interest in a \emph{normally ordered} form, whereby lowering operators ($\hat a^-_x$) occur before raising operators ($\hat a^+_x$). For example,
\begin{eqnarray}
\widehat{x^2} &=& \frac{1}{2}(\hat a_{x}^{+}+\hat a_{x}^{-})^2,\\
&= &\frac{1}{2}[(\hat a_{x}^{+})^2+(\hat a_{x}^{-})^2+2\hat a_{x}^{+}\hat a_{x}^{-}+1],
\end{eqnarray}
where we have replaced $\hat a_{x}^{-}\hat a_{x}^{+}$ by $\hat a_{x}^{+}\hat a_{x}^{-}+1$ on the second line according to our normal ordering prescription.

We do not develop these ideas any further here, but emphasize the utility of these results in the analysis of classical field calculations. Many common operators, and hence matrix elements or expectations of interest, can be expressed in terms of products of position and momentum (derivative) operators (e.g. angular momentum $\hat{L}_z=-i\left\{\hat{x}\widehat{\partial_y}-\hat{y}\widehat{\partial_x}\right\}$). These can be reformulated in terms of the normally ordered step operators that can be applied to the spectral  basis as $O(M^3)$ operations, and are exact. This avoids the $O(M^4)$ cost of transforming to quadrature grids and the associated difficulties of coming up with accurate expressions on such grids for the operators of interest.

\subsection{Evaluating the matrix elements}
The nonlinear matrix element (\ref{eq:GNL}) takes the form  given in Eq.~(\ref{EQ:GNLPW}).
 An important observation made in Ref. \cite{Dion2003a} is that these matrix elements can be computed exactly with an appropriately chosen Gauss-Hermite
quadrature. To show this we note that because the harmonic oscillator
states are of the form $\tilde\varphi_{\alpha}(\tilde x)=h_{\alpha} H_{\alpha}(\tilde x)\exp(-\tilde x^{2}/2),$
where $H_{\alpha}(\tilde x)$ is a Hermite polynomial of degree $\alpha$, the field (at any instant of time) can be written as\begin{equation}
\tilde \psi(\tilde{\mathbf{x}},\tilde t)=Q(\tilde x,\tilde y,\tilde z)e^{-(\tilde x^{2}+\tilde y^{2}+\tilde z^{2})/2},\label{eq:PsiQpoly}\end{equation}
 where 
 \begin{equation}Q(\tilde x,\tilde y,\tilde z)\equiv \sumb_{\alpha\beta\gamma}c_{\alpha\beta\gamma}(\tilde t)\,h_{\alpha}H_{\alpha}(\tilde x)h_{\beta}H_{\beta}(\tilde y)h_{\gamma}H_{\gamma}(\tilde z),
 \end{equation}is a polynomial that, as a result of the cutoff,
is of maximum degree $M-1$ in the independent variables.

Similarly, it follows that because the interaction term (\ref{eq:GNL}) is fourth
order in the field, it can be written in the form\begin{equation}
G_{\alpha\beta\gamma}=\int d^{3}\tilde{\mathbf{x}}\: e^{-2(\tilde x^{2}+\tilde y^{2}+\tilde z^{2})}P_{\alpha\beta\gamma}(\tilde x,\tilde y,\tilde z),\label{eq:FNLquad}\end{equation}
 where 
\begin{eqnarray}P_{\alpha\beta\gamma}(\tilde x,\tilde y,\tilde z)&\equiv& h_{\alpha}H_{\alpha}(\tilde x)h_{\beta}H_{\beta}(\tilde y)h_{\gamma}H_{\gamma}(\tilde z)\nonumber\\
&&\times|Q(\tilde x,\tilde y,\tilde z)|^2Q( \tilde x,\tilde y,\tilde z),
\end{eqnarray}
 is a polynomial of maximum degree $4\,(M-1)$
in the independent variables. Identifying the exponential term as
the usual weight function for Gauss-Hermite quadrature, the integral
can be exactly evaluated using a three-dimensional spatial 
grid of $8\,(M-1)^3$ points (i.e. $2\,(M-1)$ points in each direction \footnote{Since a polynomial of degree $2N-1$ is integrated exactly using a $N$ point quadrature.}), i.e.
\begin{equation}
G_{\alpha\beta\gamma} =\sum_{ijk}w_iw_jw_kP_{\alpha\beta\gamma}(\tilde{x}_i,\tilde{x}_j,\tilde{x}_k),\label{Gabc}
\end{equation}
where $\tilde x_i$ and $w_i$ are the $2\,(M-1)$ roots and weights of the 1D Gauss-Hermite quadrature with weight function $w(\tilde x)=\exp(-2\tilde x^2)$ \cite{abramowitz1964a}.

The perturbation potential can be calculated exactly on this grid if it is of the form 
\begin{equation}
\widetilde{\delta V}(\tilde{\mathbf{x}},\tilde{t})=e^{-(\tilde x^{2}+\tilde y^{2}+\tilde z^{2})}R(\tilde x,\tilde y,\tilde z),
\end{equation}
where $R(\tilde x,\tilde y,\tilde z)$ is a polynomial of maximum degree $2(M-1)$
in the independent variables. However, if we assume the perturbation is small, then it will be permissible to evaluate it approximately on the same quadrature grid used for the nonlinear term.

\subsection{Overview of numerical procedure}\label{SEC:HARMnummeth}
Here we briefly overview how the quadrature described  above can be efficiently implemented numerically. This differs slightly from the planewave case by virtue of the less trivial nature of the quadrature weights and weight functions. Starting from the basis set representation of the field (i.e. $\{c_{\alpha\beta\gamma}\}$) at an instant of time $\tilde t$, the steps for calculating the matrix elements are as follows:
\begin{enumerate}
\item We transform the field to a spatial representation  according to 
\begin{equation}
\tilde{\psi}(\tilde{\mathbf{x}}_{ijk},\tilde t)=\sumb_{\alpha\beta\gamma}U_{i\alpha}U_{j\beta}U_{k\gamma}\,c_{\alpha\beta\gamma}(\tilde t),\label{Harm2pos}
\end{equation}
where the transformation matrices are defined as the 1D basis states evaluated on the quadrature grid, i.e.
\begin{equation}
U_{i\alpha}=\tilde\varphi_{\alpha}(\tilde{x}_i).\label{harmU}
\end{equation}

\item The integrands of the matrix elements (\ref{eq:Fpert}) and (\ref{eq:GNL})  are then constructed as quadratures by appropriately dividing by the weight function and pre-multiplying by the weights \footnote{Here we form $e^{2|\tilde{\mathbf{x}}_{ijk}|^2}|\tilde{\psi}|^{2}\tilde{\psi}$ as this corresponds to the polynomial  ($P$) required for the quadrature (see Eq.~(\ref{Gabc})).}, i.e.
\begin{eqnarray}
f(\tilde{\mathbf{x}}_{ijk})&\equiv&w_iw_jw_ke^{2|\tilde{\mathbf{x}}_{ijk}|^2}\widetilde{\delta V}(\tilde{\mathbf{x}}_{ijk},\tilde{t})\tilde{\psi}(\tilde{\mathbf{x}}_{ijk},\tilde{t}),\\
g(\tilde{\mathbf{x}}_{ijk})&\equiv&w_iw_jw_ke^{2|\tilde{\mathbf{x}}_{ijk}|^2}|\tilde{\psi}(\tilde{\mathbf{x}}_{ijk},\tilde{t})|^{2}\tilde{\psi}(\tilde{\mathbf{x}}_{ijk},\tilde{t}).\label{Eqgtrans}
\end{eqnarray}

\item Inverse transforming these integrand functions yields the desired matrix elements:
\begin{eqnarray}
F_{\alpha\beta\gamma} &=& \sum_{ijk}U_{i\alpha}^*U_{j\beta}^*U_{k\gamma}^*f(\tilde{\mathbf{x}}_{ijk}),\label{harmFtrans}\\
G_{\alpha\beta\gamma} &=& \sum_{ijk}U_{i\alpha}^*U_{j\beta}^*U_{k\gamma}^*g(\tilde{\mathbf{x}}_{ijk}).
\end{eqnarray}
\end{enumerate}
The slowest step in this procedure is carrying out the basis transformation. The computational cost of this step is $O(M^4)$ when carried out as a series of matrix multiplications.  

\subsection{Other transforms}\label{SEC:HARMtranforms}
For completeness we describe how the basis transformations can be generalized for use in preparing states in the spectral basis and transforming to position and momentum grids. While being unrelated to  propagation, these procedures are part of an essential set of tools for the general application of the method and analysis of results.

\subsubsection{Projecting a position space state onto the spectral basis}\label{SEC:Harmposproj}
Frequently we are presented with a field specified in the position basis, i.e. $\tilde{\psi}^A(\tilde{\mathbf{x}})$, and need to obtain its spectral representation, i.e.
\begin{equation}
c_{\alpha\beta\gamma}^A = \int d^{3}\tilde{\mathbf{x}}\,\tilde{\varphi}_{\alpha}^{*}(\tilde{x}) \tilde{\varphi}_{\beta}^{*}(\tilde{y}) \tilde{\varphi}_{\gamma}^{*}(\tilde{z}) \tilde{\psi}^A(\tilde{\mathbf{x}}). \label{cposproj}
\end{equation}
If the modes of the  classical region span the function $\tilde{\psi}^A(\tilde{\mathbf{x}})$ then this procedure provides an exact representation, however, in general the resulting coefficients  $c_{\alpha\beta\gamma}^A$ represent the function projected into the classical region, i.e., $\PC\{\tilde{\psi}^A\}$. 

In practice we implement this transform onto the spectral basis in a similar manner to that used in Eq.~(\ref{harmFtrans}), i.e., as
\begin{equation}
c_{\alpha\beta\gamma}^A= \sum_{ijk}U_{i\alpha}^*U_{j\beta}^*U_{k\gamma}^*\,w_iw_jw_k\,\tilde{\psi}^A(\tilde{\mathbf{x}}_{ijk}),\label{tranfcA}
\end{equation}
where the transformation matrices are defined as the 1D basis states evaluated on the quadrature grid, i.e.,
$U_{i\alpha}=\tilde\varphi_{\alpha}(\tilde{x}_i)$ as in Eq.~(\ref{harmU}
but now the quadrature is different to that used to evaluate the nonlinear matrix elements. Replacing $\tilde{\psi}^A$ by $\PC\{\tilde{\psi}^A\}$ in Eq.~(\ref{cposproj}) we note that integrand is of the form of a exponential $\exp(-|\tilde{\mathbf{x}}|^2)$ times a polynomial function of maximum degree $2\,(M-1)$ in each of the coordinates. These integrals are exactly computed in each direction using Gauss-Hermite quadrature defined for the weight function $w(\tilde x)=\exp(-\tilde x^2)$   with $M$ roots $\tilde x_i$ and weights $w_i$.

\subsubsection{Transforming the classical field to arbitrary position and momentum grids}
An important aspect of any classical field method is the ability to transform the results to desired grids for analysis and/or visualization. This is particularly convenient for the spectral representation we have adopted here. The basic procedure is as indicated in Eq.~(\ref{Harm2pos}) except that the grid points are now arbitrary and need not be related to any quadrature grid. Taking one direction to have a single point, e.g., $\{\tilde{x}_j\}=\tilde{x}_1$,  allows us to take slices of the classical field.

As shown in Eq.~(\ref{EQmtmfield}) in the spectral basis there is a simple relationship between the position and momentum representations. Thus to obtain the momentum space field we  use   $U_{i\alpha}^p=(-i)^{\alpha}\tilde\varphi_{\alpha}(\tilde{p}_i)$ as the transformation matrices, i.e. 
\begin{equation}
\tilde{\Phi}(\tilde{\mathbf{p}}_{ijk},\tilde t)=\sumb_{\alpha\beta\gamma}U_{i\alpha}^pU_{j\beta}^pU_{k\gamma}^p\,c_{\alpha\beta\gamma}(\tilde t).\label{Harm2mtm}
\end{equation}

\subsubsection{Column densities}
Computing column densities is  important for modeling ultra-cold Bose gases, since this is what is measured in experiments using absorption imaging (the primary method of analyzing these systems). For example, when the \emph{in situ} system is imaged using light propagating along the $z$-direction, the observable corresponds to the column density defined as
\begin{equation}
\tilde{n}_c(\tilde{x},\tilde{y}) \equiv \int d\tilde{z}\,|\tilde{\psi}(\tilde{\mathbf{x}})|^2.\label{coldenz}
\end{equation}
On the other hand, often the trapping potential is turned off and the system is allowed to expand freely before it is imaged.
If the expansion time is sufficiently long the measured signal is related to a momentum space column density, e.g. for imaging along $\tilde z$ 
\begin{equation}
\tilde{n}_c(\tilde{p}_x,\tilde{p}_y) \equiv \int d\tilde{p}_z\,|\tilde{\Phi}(\tilde{\mathbf{p}})|^2.
\end{equation}
If the imaging direction (i.e. direction along which we integrate) corresponds to a coordinate direction the column density can be conveniently and exactly evaluated. We will consider the case in Eq.~(\ref{coldenz}) for definiteness, though the same procedure immediately applies to the momentum case, and for column densities taken along other axes.

We transform the field to a position grid $\tilde{\psi}(\tilde{\mathbf{x}}_{ijk})$, where the $\tilde x$ and $\tilde y$ grids can be chosen arbitrarily, but the $\tilde z$ grid needs to be a $M$ point quadrature grid of the type discussed below Eq.~(\ref{tranfcA}).  This choice for the $\tilde z$ grid ensures that 
\begin{equation}
\tilde{n}_c(\tilde{x}_i,\tilde{y}_j)=\sum_kw_ke^{\tilde{z}_k^2}|\tilde{\psi}(\tilde{\mathbf{x}}_{ijk})|^2,
\end{equation}
exactly computes the $\tilde z$ integral in Eq.~(\ref{coldenz}) where the weights are those described below Eq.~(\ref{tranfcA}).
 
\subsection{Three body term}
 Another important term that often needs to be evaluated (e.g.~see Ref.~\cite{Wuster2007a}) is the so called three-body term, which can be implemented by adding the term $-\frac{1}{2}i\tilde{K}_{3}|\tilde\psi|^4\tilde\psi$ to the right hand side of Eq.~(\ref{eq:numGPE1}), where   $\tilde{K}_{3}$ characterizes the loss of atoms due to three-body recombination. Generally this process is small in comparison to the two-body interaction term (of strength $C$), and causes the field to lose normalization.
 
Including this term in the evolution equation (\ref{eq:GPEshobasis}) we obtain
 \begin{equation}
\frac{\partial c_{n}}{\partial\tilde t}  =  -i\left[\tilde\epsilon_{n}c_n+F_n+CG_{n}\right]-\frac{1}{2}{\tilde{K}_3}J_n,\label{eq:GPEshobasis3B}
 \end{equation} 
 where  we have introduced the matrix element
 \begin{equation}
 J_{n}=\int d^3\tilde{\mathbf{x}}\,\tilde{\phi}_n^*(\tilde{\mathbf{x}})|\tilde{\psi}(\tilde{\mathbf{x}},\tilde{t})|^4\tilde{\psi}(\tilde{\mathbf{x}},\tilde{t}).
 \end{equation}
 Decomposing this into 1D eigenstate basis (i.e. $n\to\{\alpha\beta\gamma\}$) we have that
 \begin{equation}
 J_{\alpha\beta\gamma}=\int d^3\tilde{\mathbf{x}}\,e^{-3(\tilde{x}^2+\tilde{y}^2+\tilde{z}^2)}S_{\alpha\beta\gamma}(\tilde{x},\tilde{y},\tilde{z}),
 \end{equation}
 where $S_{\alpha\beta\gamma}$ is a polynomial of maximum degree $6(M-1)$ in the independent variables. 
 
Using the same arguments as applied to the normal interaction term (see Sec.~\ref{oscimatelem}), we can show that this term could be evaluated exactly on a three-dimensional spatial grid consisting of $27(M-1)^3$ points (i.e~$3(M-1)$ points along each direction) using a quadrature appropriate  to the weight function $w(\tilde x)=\exp(-3\tilde{x}^2)$ in each direction.
 
 However, like the perturbative potential, usually this term has a rather small effect on the dynamics and the additional expense of implementing a special transform to exactly evaluate the $J$-matrix elements is unnecessary. In such cases it should be an acceptable approximation to calculate these matrix elements on the same grids used for the two-body interaction.

\subsection{Convergence and future numerical work}  
 \begin{table}[htbp]
   \centering 
\begin{tabular}{l  c  c  c }
\hline\hline
Relative Tolerance & Number of steps & $\delta N$ & $\delta X$ \\
\hline
 $ 1\times10^{-2} $ & $131$ & $2.54\times10^{-2}$ & $3.13\times10^{-1}$\\ 
 $ 5\times10^{-3}$ & $135$ &  $2.04\times10^{-2}$  & $1.94\times10^{-1}$\\ 
 $ 1\times10^{-3}$ & $186$ &  $6.72\times10^{-3}$  & $2.01\times10^{-2}$\\ 
 $ 5\times10^{-4}$ & $213$ &  $3.52\times10^{-3}$  & $5.39\times10^{-3}$\\ 
 $ 1\times10^{-4}$ & $294$ & $7.13\times10^{-4}$  & $2.25\times10^{-4}$\\ 
 $ 5\times10^{-5}$ & $335$ &  $3.57\times10^{-4}$  & $5.54\times10^{-5}$\\ 
 $ 1\times10^{-5}$ & $455$ &  $7.02\times10^{-5}$  & $2.15\times10^{-6}$\\ 
 $ 5\times10^{-6}$ & $527$ &  $3.50\times10^{-5}$  & $5.29\times10^{-7}$\\ 
  $1\times10^{-6}$ & $725$ &  $6.93\times10^{-6}$  & $2.05\times10^{-8}$\\ 
  $5\times10^{-7}$ & $829$ &  $3.46\times10^{-6}$  & $4.98\times10^{-9}$\\ 
 $ 1\times10^{-7}$ & $1139$ &  $6.92\times10^{-7}$  & $1.67\times10^{-10}$\\ 
 $ 5\times10^{-8}$ & $1311$ &  $3.45\times10^{-7}$  & $3.28\times10^{-11}$\\ 
 $ 1\times10^{-8}$ & $1797$ &  $6.92\times10^{-8}$  & N/A\\   \hline\hline
\end{tabular}
\caption{\label{convtable} Convergence properties of evolution algorithm. The relative error tolerance, number of steps needed to obtain that error tolerance, and the final error measures $\delta N$ and $\delta X$ are shown (see text). Simulation with $10^{-8}$ tolerance is used as $c^A_j$ in measuring error of less accurate calculations. Other parameters: $\tilde{t}=2\pi$, $C=1000$, $\lambda_x=4$, $\lambda_y=1$, $\tilde{\epsilon}_{\rm{cut}}=35$ and the initial state has energy $\tilde{E}=18$ (see Sec.~\ref{Results}). }
\end{table}  

Here we present some evolution convergence results for our algorithm and discuss avenues for future development of PGPE numerics. 
Unlike the zero temperature Gross-Pitaevskii equation, where convergence can be ensured by increasing the basis size, here such a procedure will increase the classical region size (number of degrees of freedom) and hence effect the thermal properties of the system \footnote{We note that such convergence properties for finding the zero temperature ground state of the time-independent Gross-Pitaevskii equation using an oscillator-state based spectral representation, like what we use here for the PGPE, has been studied previously, e.g. see Ref.~\cite{Tiwari2006}).}. Thus for the PGPE formalism the essential requirement is that all modes within the classical region are propagated precisely.

\begin{figure}[htbp] 
   \centering
   \includegraphics[width=3.2in]{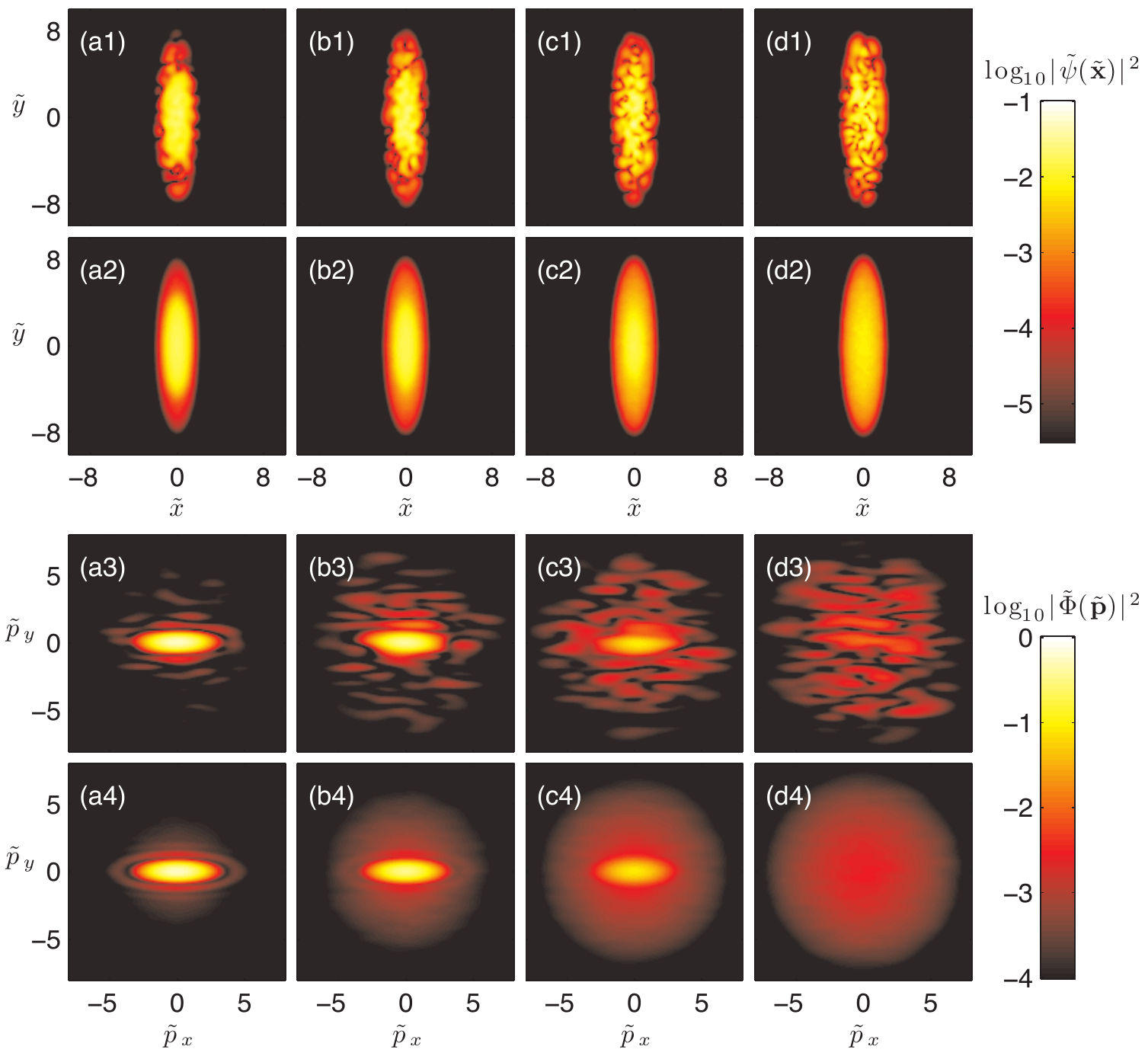} 
   \caption{(color online) Position and momentum density slices in the $xy$-plane. Row (a1)-(d1) instantaneous position density slices for $\tilde z=0$. 
    Row (a2)-(d2) time-averaged position density slices for $\tilde z=0$.    Row (a3)-(d3) instantaneous momentum density slices for $\tilde{p}_z=0$.  Row (a4)-(d4) time-averaged momentum density slices for $\tilde{p}_z=0$. Parameters: Column (a1)-(a4) $\tilde{E}=11.64$, $f_{\rm{cond}}=0.87$, column (b1)-(b4)  $\tilde E= 15.07$, $f_{\rm{cond}}=0.61$, column (c1)-(c4)  $\tilde E= 20.70$, $f_{\rm{cond}}=0.24$, column (d1)-(d4)  $\tilde E=  25.83$, $f_{\rm{cond}}=0.01$. Other parameters: $C=750$, $\lambda_x=4$, $\lambda_y=1$ and $\tilde{\epsilon}_{\rm{cut}}=35$. 
   }
   \label{fig:denXY}
\end{figure}

\begin{figure}[htbp] 
   \centering
   \includegraphics[width=3.2in]{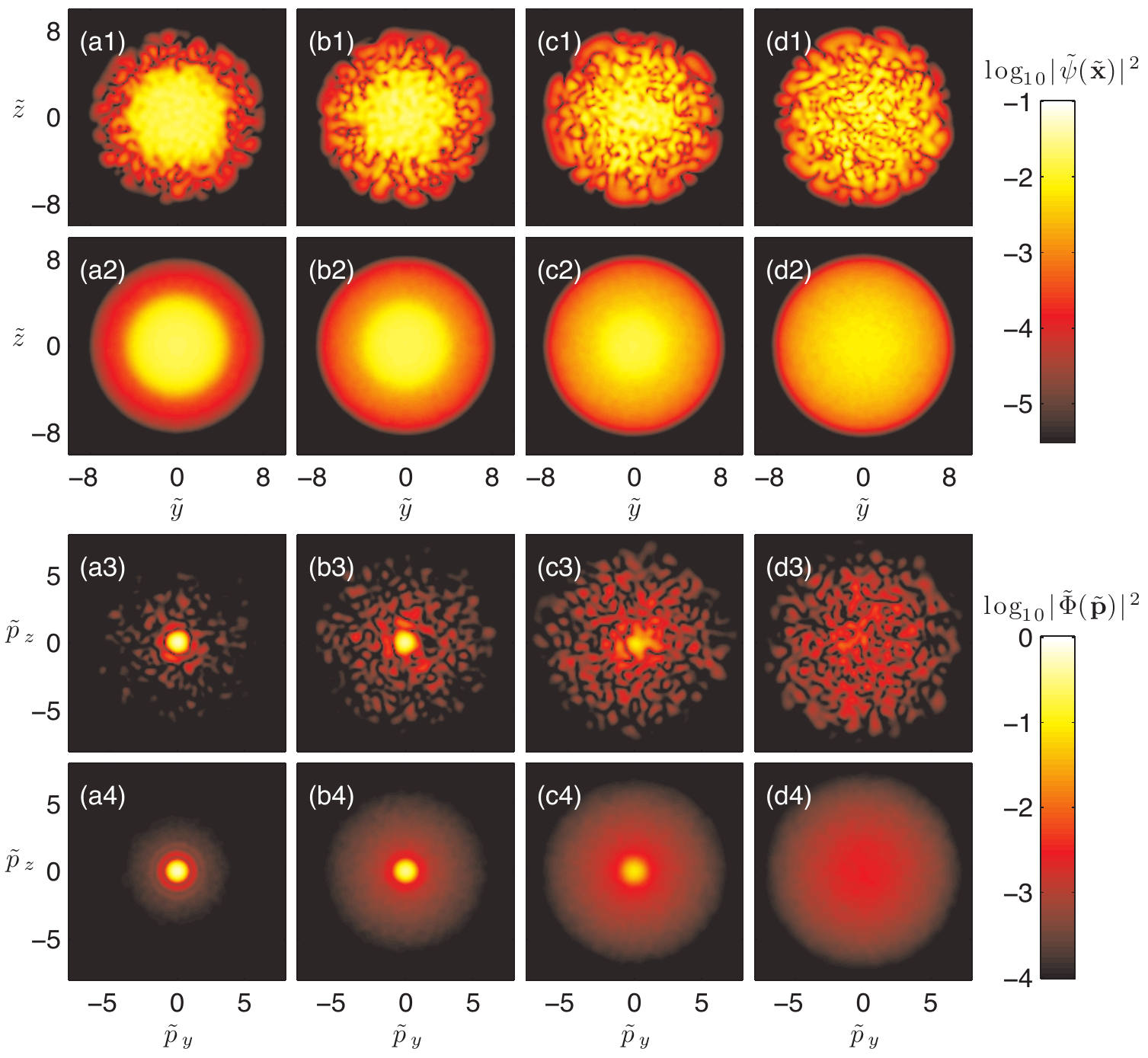} 
   \caption{(color online) Position and momentum density slices in the $yz$-plane. Row (a1)-(d1) instantaneous position density slices for $\tilde x=0$. 
    Row (a2)-(d2) time-averaged position density slices for $\tilde x=0$.    Row (a3)-(d3) instantaneous momentum density slices for $\tilde{p}_x=0$.  Row (a4)-(d4) time-averaged momentum density slices for $\tilde{p}_x=0$. Parameters: Column (a1)-(a4) $\tilde{E}=11.64$, $f_{\rm{cond}}=0.87$, column (b1)-(b4)  $\tilde E= 15.07$, $f_{\rm{cond}}=0.61$, column (c1)-(c4)  $\tilde E= 20.70$, $f_{\rm{cond}}=0.24$, column (d1)-(d4)  $\tilde E=  25.83$, $f_{\rm{cond}}=0.01$. Other parameters: $C=750$, $\lambda_x=4$, $\lambda_y=1$ and $\tilde{\epsilon}_{\rm{cut}}=35$. 
    }
   \label{fig:denYZ}
\end{figure}

\begin{figure}[htbp] 
   \centering
   \includegraphics[width=3.2in]{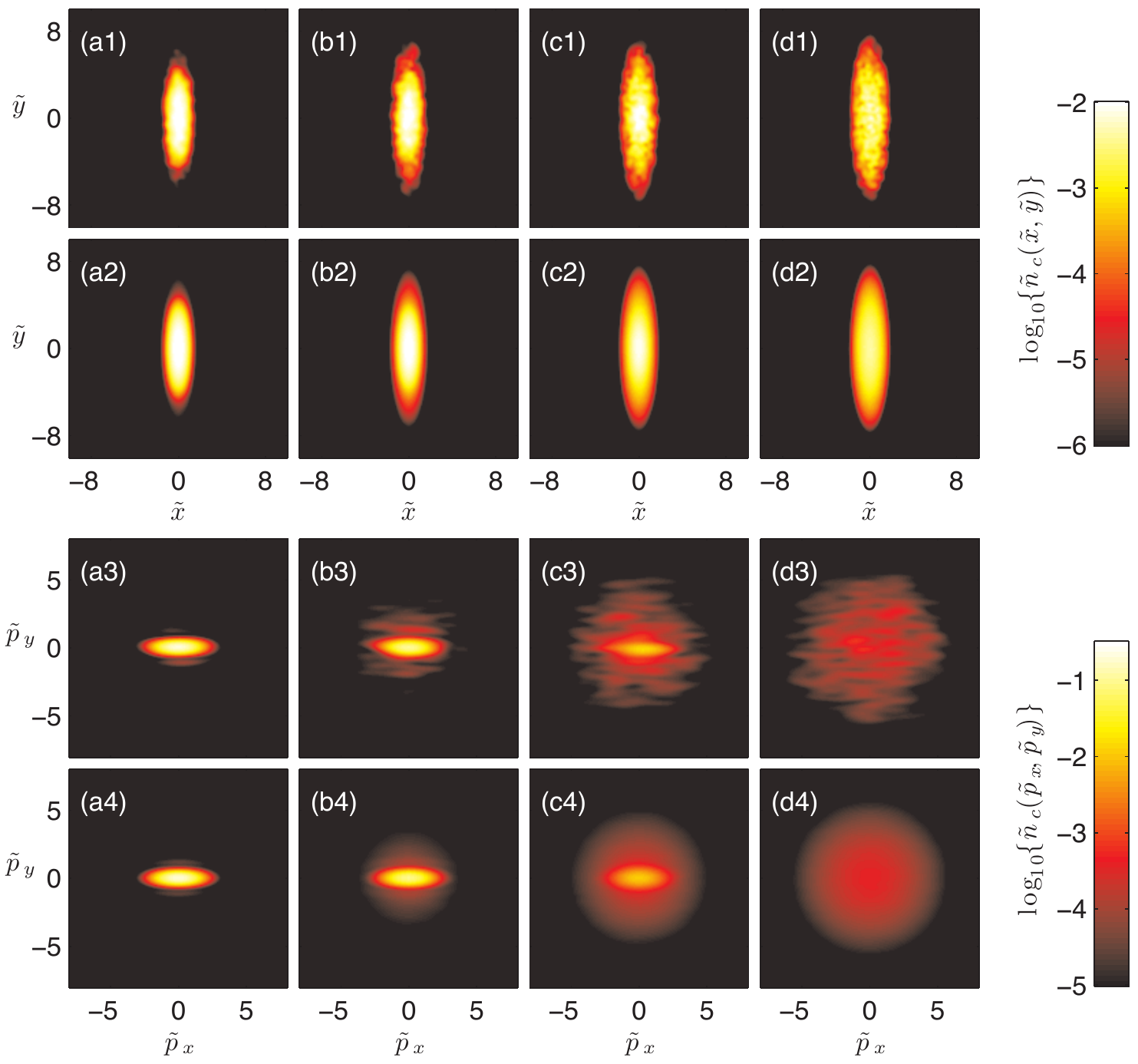} 
   \caption{(color online) Position and momentum column densities integrated along the $\tilde z$-axis. Row (a1)-(d1) instantaneous position column densities. 
    Row (a2)-(d2) time-averaged position column densities.    Row (a3)-(d3) instantaneous momentum column densities.  Row (a4)-(d4) time-averaged momentum column densities. Parameters: Column (a1)-(a4) $\tilde{E}=11.64$, $f_{\rm{cond}}=0.87$, column (b1)-(b4)  $\tilde E= 15.07$, $f_{\rm{cond}}=0.61$, column (c1)-(c4)  $\tilde E= 20.70$, $f_{\rm{cond}}=0.24$, column (d1)-(d4)  $\tilde E=  25.83$, $f_{\rm{cond}}=0.01$. Other parameters: $C=750$, $\lambda_x=4$, $\lambda_y=1$ and $\tilde{\epsilon}_{\rm{cut}}=35$. 
    }
   \label{fig:denCXY}
\end{figure}

We have used an adaptive step Runge-Kutta-Fehlberg algorithm to evolve the classical field equation (\ref{eq:GPEshobasis}) with a specified relative error tolerance. 
Since computing the matrix elements for our harmonically trapped algorithm is of computational cost $O(M^4)$ the development of higher order or more efficient propagation algorithms would be desirable (e.g.~see Refs.~\cite{Muruganandam2003,Adhikari2002,Xu2006,Chin2007} for various algorithms developed for evolving the zero temperature Gross-Pitaevskii equation). 
For example, the simulation presented in column (d) of Figs.~\ref{fig:denXY}-\ref{fig:denCXY} has a basis of $M_T=1857$ modes propagated over a time interval of $\tilde{t}\sim200\pi$.
At a relative tolerance of $10^{-6}$ this simulation took approximately 2hrs and 23 minutes on a 2.66GHz Xeon (Woodcrest) processor, consisting of 61,147 steps including 3,013 failed steps. In Table \ref{convtable} we examine the evolution algorithm convergence properties using the two measures
\begin{eqnarray}
\delta X &=& \sum_{j=1}^{M_T}|c_j(\tilde{t})-c^A_j(\tilde{t})|^2,\\
\delta N &=& 1- \sum_{j=1}^{M_T}|c_j(\tilde{t})|^2,
\end{eqnarray}
i.e.~a difference measure of the final states and the loss in normalization, where $c^A_j(\tilde{t})$ are the mode amplitudes at time $\tilde{t}$ of a more accurate simulation.

While we have strived to ensure highly accurate time propagation, statistical mechanical arguments suggest that this may be unnecessary and some additional \emph{noise} from imprecise evolution may assist the system explore the ensemble of available states more rapidly, as long as this noise does not effect the constants of motion describing the system. For our evolution algorithm the normalization of the field is not conserved and anecdotal evidence suggests that monitoring changes in normalization can be used as a summative assessment of the evolution accuracy, e.g.~for the simulation described above 
the final state normalization was 0.99989.

Here we have considered a spectral cutoff in the \emph{ideal} single particle basis, \emph{i.e}.~harmonic oscillator states. (We refer readers to Ref.~\cite{Bradley2008a} for a discussion of how numerics can be implemented for a rotating trapped system). For sufficiently high energy cutoff, this basis approximately diagonalizes the many-body problem. However, for situations requiring lower energy cutoffs (e.g. when $T\ll T_c$), the Hartree-Fock interacting modes would form a better basis for defining the classical region. Since these modes do not have an associated quadrature, the scheme presented here cannot be immediately extended to this regime.  Another pathway, which may allow more flexibility in the nature of the cutoff, would be to use a more efficient representation (e.g.~plane-wave/Fourier transform) and approximately implement the projection.

\section{Results}\label{Results}


In this section we consider the application of the PGPE formalism to a harmonically trapped system with a fixed energy cutoff, and take $\widetilde{\delta V}=0$. To elucidate features of our method we consider an anisotropic harmonic trap with $\lambda_x=4$ and $\lambda_y=1$, i.e. the system has a \emph{fat pancake} geometry with the $x$-direction being tightly confined. Taking $\tilde{\epsilon}_{\rm{cut}}=35$ there are  $M_T=1857$ single particle modes in the classical region with the maximum order of eigenstate occurring in each coordinate direction being  $(\alpha_{\max})_x=8$, $(\alpha_{\max})_y=32$, and $(\alpha_{\max})_z=32$.

The PGPE is ergodic and the generic method to study finite temperature regimes is to begin with a randomized initial state with some definite energy, $\tilde E$, as quantified by the energy functional
\begin{equation}
\tilde{E}[\tilde{\psi}] = \int d^3\tilde{\mathbf{x}}\,\tilde{\psi}^*\left\{ \tilde H_{\rm sp}+ \frac{C}{2} |\tilde\psi|^{2}\right\}\tilde\psi.\label{eq:numEfunc} 
\end{equation}
This energy is a constant of motion for the PGPE (\ref{eq:numGPE1}) and forms a convenient macroscopic constraint for specifying the thermal state of the system. The procedure for making such energy states is rather arbitrary. We choose to make use of the Thomas-Fermi approximation to the ground state of the energy functional, given by
\begin{equation}
\tilde{\psi}_{\rm TF}(\tilde{\mathbf{x}})=\sqrt{\frac{\mu_{\rm TF}-\tilde{V}_{\rm trap}(\tilde{\mathbf{x}})}{C}}\,\,\theta\left(\mu_{\rm TF}-\tilde{V}_{\rm trap}(\tilde{\mathbf{x}})\right),
\end{equation}
where $\theta(x)$ is the unit step function and $\mu_{\rm TF}=\frac{1}{2}(15\lambda_x\lambda_yC/4\pi)^{2/5}$ is the Thomas-Fermi chemical potential \cite{Dalfovo1999}. 
We mix the Thomas-Fermi state (projected onto the spectral basis according to the procedure given in Sec.~\ref{SEC:Harmposproj}) with a 
\emph{high energy} randomized state \footnote{We construct a random vector according to $c_n=a_n+ib_n$ where $a_n$ and $b_n$ are normally distributed random values. We then normalize this state so that $\sum_n|c_n|^2=1$.} in the appropriate ratio to obtain a normalized state of the desired energy. Such states, for a range of energies, are propagated according to the PGPE for a time period of $\tilde t=200\pi$ and $500$ states of the field are saved at equally spaced times during the evolution, i.e. at times $\tilde{t}_j=2\pi j/5$, with $0\le j\le500$.

Due to the stochastic nature of the initial condition for the simulation the instantaneous behavior of the system is of reduced importance, instead the quantities of interest are the macroscopic observables that can be computed from the system evolution. We consider a few such observables here, and refer the reader to Ref.~\cite{Blakie2005a} for a more general discussion on this topic.

One observable we will concern ourselves with is the average position density calculated as the time-average
\begin{equation}
\left\langle |\tilde{\psi}(\tilde{\mathbf{x}})|^2\right\rangle = \frac{1}{M_{\rm s}}\sum_j|\tilde{\psi}(\tilde{\mathbf{x}},\tilde{t}_j)|^2,\label{psiTA}
\end{equation}
 where the summation is taken over some subset of $M_{\rm s}$ saved states. Here we use the last $350$ of the saved states for averaging. We use the first part of the evolution (i.e. up to $\tilde t=60\pi$) to allow the arbitrarily chosen initial state to relax to equilibrium, a point we consider further in Sec.~\ref{SEC:Relax}. Similarly we can construct the momentum density by time-averaging the Fourier transformed classical field (\ref{EQmtmfield}).
 
Another quantity of interest is the condensate fraction. The definition we use was provided by Penrose and Onsager \cite{Penrose1956a}, and identifies the condensate fraction $f_{\rm{cond}}$ as the largest eigenvalue of the one-body density matrix, defined in terms of the field as $G^{1B}(\tilde{\mathbf{x}},\tilde{\mathbf{x}}^\prime)=\langle \tilde \psi^*(\tilde{\mathbf{x}})\tilde \psi(\tilde{\mathbf{x}}^\prime)\rangle$. In our formalism this quantity is equivalently and much more efficiently computed in the mode basis as
 \begin{equation}
 G_{mn}^{1B}=\langle c^*_mc_n\rangle,
 \end{equation}
 where  $\langle\rangle$ is evaluated using time-averaging [e.g., see Eq.~\ref{psiTA}]. It is also possible to calculate thermal parameters from the dynamical evolution of the field, such as temperature and chemical potential, and we refer the reader to Ref.~\cite{Davis2005a} for more details.
  
 \subsection{Instantaneous and time averaged density profiles} 
 Our results for the instantaneous and averaged density profiles are shown in Figs.~\ref{fig:denXY}  and \ref{fig:denYZ}.
 In Fig.~\ref{fig:denXY} the system is viewed in the $xy$-plane using a $\tilde z = 0$ slice through the field. 
 The position space density clearly reveals the  anisotropy of the confining potential. As the energy of the system increases it fluctuates more strongly (as revealed in the instantaneous density slices), however the averaged density profiles change quite gradually. 
 In contrast, the momentum space density (shown as $\tilde{p}_z=0$ slices) changes much more significantly with system energy. At low energies [e.g., Fig.~\ref{fig:denXY}(a3) and (a4)] the condensate is prominent in the system, and is easily identified as the dense anisotropic momentum peak centered at zero momentum. This momentum anisotropy is the opposite of that observed in position space due to the Heisenberg relationship between position and momentum size of the  condensate wavepacket along each direction. We also see a background radially symmetric distribution in the averaged momentum density [e.g., Fig.~\ref{fig:denXY}(a4)] that increases in prominence (at the expense of the condensate) as the energy of the system increases [e.g., see figures along the row Fig.~\ref{fig:denXY}(a4)-(d4)]. This background constitutes what is usually referred to as the \emph{thermal cloud}  of the system -- or at least the component of this that exists within the classical region \footnote{There are more such modes in the incoherent region which is not described by the PGPE.}. Comparing  the averaged and respective instantaneous momentum density profiles [i.e., see Fig.~\ref{fig:denXY} along the rows (instantaneous) (a3)-(d3) and (averaged) (a4)-(d4)] we see that the modes contributing to this radially symmetric density fluctuate quite strongly, and adequate time-averaging to obtain their averaged profiles is essential.

In Fig.~\ref{fig:denYZ} the system is viewed in the  $yz$-plane  using  a $\tilde x=0$ slice through the field
($\tilde{p}_x$=0 slice for the momentum space results). 
The averaged position and momentum  space density are isotropic like the trapping potential in this plane. 
These results also allow us to appreciate the role of the classical region cutoff on typical fluctuations in the system. Compare the instantaneous momentum densities in Fig.~\ref{fig:denXY}(d3) and Fig.~\ref{fig:denYZ}(d3). In Fig.~\ref{fig:denYZ}(d3) the length scale for fluctuations is the same in the $\tilde{p}_y$ and $\tilde{p}_z$ directions, reflecting the symmetry of the potential. In Fig.~\ref{fig:denXY}(d3) the length scale for fluctuations is much longer in the $\tilde{p}_x$ direction than in the $\tilde{p}_y$ direction. We do not see the same features in the position space fluctuations, e.g. in  Fig.~\ref{fig:denXY}(d1) and Fig.~\ref{fig:denYZ}(d1) where the length scales in all directions appear to be similar.  These observations can be explained by a simple argument based on a semiclassical analysis of the single particle Hamiltonian (\ref{eq:H0harm}): The maximum position extent of the system in the $j$-direction is $\tilde{X}_j\sim\sqrt{2\tilde{\epsilon}_{\rm cut}}/\lambda_j$, and using that the number of modes in this direction is $M_j\sim\tilde{\epsilon}_{\rm cut}/\lambda_j$, we obtain that the typical fluctuation length scale is $\Delta \tilde{x}_j\sim\sqrt{2/\tilde{\epsilon}_{\rm cut}}$ 
\footnote{Here we have taken $\Delta\tilde{x}_j\sim\tilde{X}_j/M_j$ and similarly $\Delta\tilde{p}_j\sim\tilde{P}_j/M_j$.}, i.e., independent of trap frequency (i.e. $\lambda_j$) in position space. In momentum space we instead have that the momentum extent of the field,  $\tilde{P}_j\sim\sqrt{2\tilde{\epsilon}_{\rm cut}}$, is independent of the trap frequency, while the length scale for fluctuations $\Delta\tilde{p}_j\sim \lambda_j\sqrt{2/\tilde{\epsilon}_{\rm cut}}$ is proportional to the trap frequency. This qualitatively agrees with the observation in Fig.~\ref{fig:denXY}(d3) that the momentum fluctuation length scale is about 4 times longer in the $\tilde{p}_x$ direction than in the $\tilde{p}_y$ direction.

For comparison with the density slices we have just discussed in Fig.~\ref{fig:denCXY}  we show the instantaneous and time-averaged column densities in position and momentum space. Taking the column density in some sense averages the fluctuations in the instantaneous system state compared to the density slice.  
The instantaneous column densities closely approximate the images taken in experiment, and thus the fluctuations seen should be measurable, however we note that our results here exclude the contribution from the incoherent region which in general will be important to include.

\subsection{Relaxation to equilibrium}\label{SEC:Relax}
\begin{figure}[htbp] 
   \centering
   \includegraphics[width=3.2in]{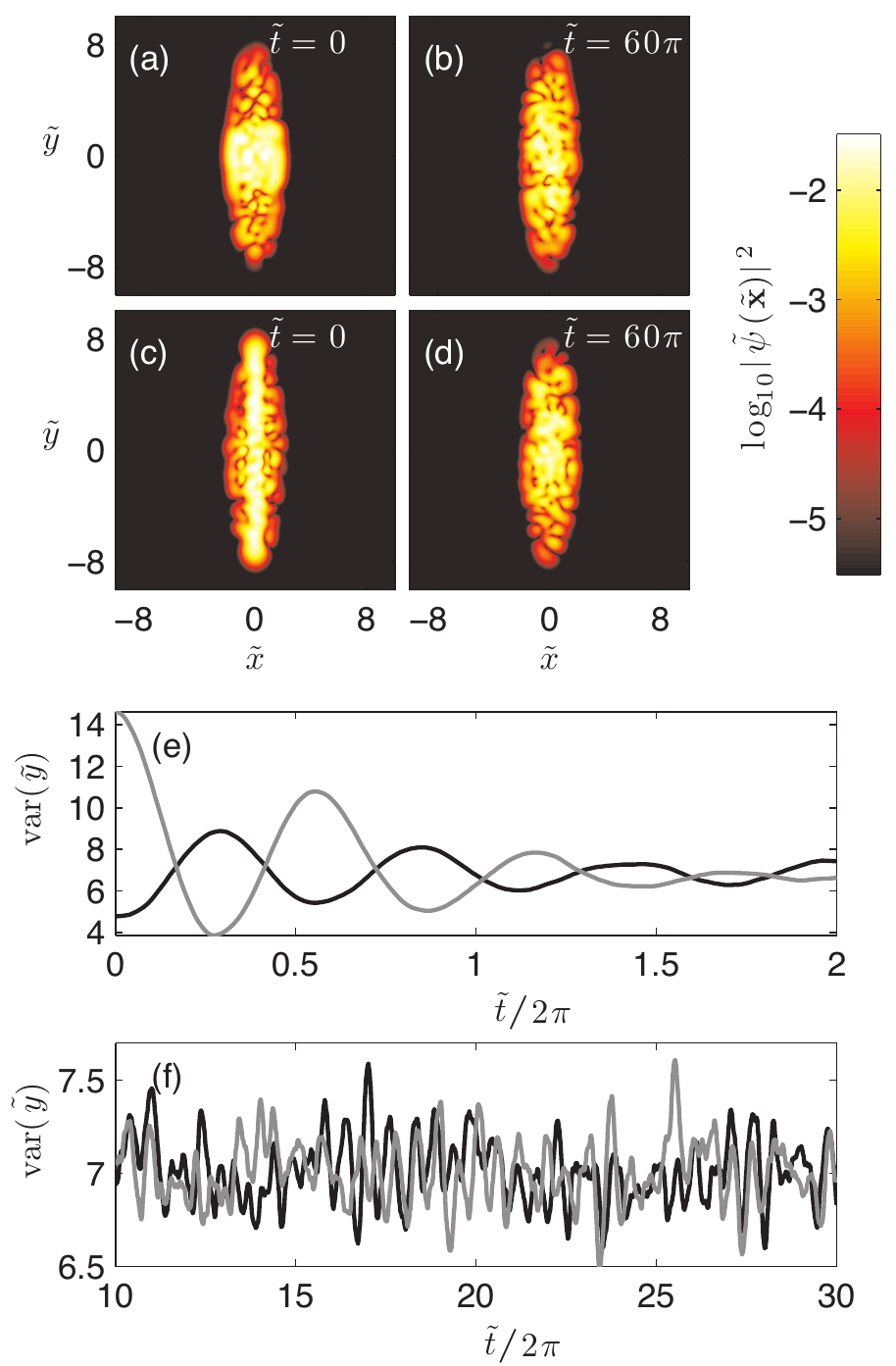} 
   \caption{(color online) Relaxation to equilibrium. Density slices of the two non-equilibrium initial states (a) and (c), and the respective states they evolve to at $\tilde{t}=60\pi$ (b) and (d). Both states have $\tilde{E}=22.00$.  The variance of the classical field in the $\tilde{y}$-direction (e) during the first few trap periods and (f) after 10 trap periods. Result for initial state (a) (black line) and initial state (b) (grey line).  Other parameters: $C=750$, $\lambda_x=4$, $\lambda_y=1$ and $\tilde{\epsilon}_{\rm{cut}}=35$. 
    }
   \label{fig:denNEQ}
\end{figure}

Finally we consider some dynamical aspects of how a \emph{distorted} system, described by the PGPE, returns to an equilibrium state.  To do this we construct two initial states with the same total energy but rather different density distributions. Density slices of these two states are shown in Figs.~\ref{fig:denNEQ}(a) and (c). Our preparation procedure for those two states is as described before, except that here we choose to use a Thomas-Fermi state that is distorted so as to be extended along the $\tilde x$ and $\tilde y$ directions, respectively. As conjectured earlier, since both states have the same energy they should relax to the same (dynamical) equilibrium. In Figs.~\ref{fig:denNEQ}(b) and (d) we show the density slices of these states after propagating the system for $30$ ($\tilde z$-direction) trap periods. The density slices of these final states qualitatively look much more similar to each other than the initial states did, consistent with both systems reaching the same equilibrium. However, we can provide other evidence for the system returning to equilibrium. In Figs.~\ref{fig:denNEQ}(e) and (f) we examine the position variance of the classical field along the $\tilde y$ direction defined as
\begin{equation}
\rm{var}\left(\tilde{y}(\tilde{t})\right) \equiv\left\langle \tilde{y}(\tilde{t})^2\right\rangle-\left\langle\tilde{y}(\tilde{t})\right\rangle^2,
\end{equation}
i.e.~the quantum mechanical average of the field at time $\tilde{t}$.
This provides a measure of the system width in that direction.  We note that this quantity is efficiently evaluated with the step operator formalism discussed in Sec.~\ref{SEC:stepops}.
Fig.~\ref{fig:denNEQ}(e) reveals that $\rm{var}\left(\tilde{y}(\tilde{t})\right)$ initially shows strong evolution that is characteristic of the initial state being far from equilibrium. This initial evolution is strikingly different for the two cases reflecting that  initial states considered here are respectively compressed [Fig.~\ref{fig:denNEQ}(a)] and extended [Fig.~\ref{fig:denNEQ}(c)] in the $\tilde y$ direction. After a few trap periods the dynamics of $\rm{var}(\tilde y)$ is seen to reduce considerably.

In Fig.~\ref{fig:denNEQ}(f) we examine dynamics of $\rm{var}(\tilde y)$ after the system has relaxed for $10$ trap periods. The fluctuations in $\rm{var}(\tilde y)$ are much reduced in amplitude compared to what was observed in the initial evolution. Indeed, the fluctuations here are consistent with thermal fluctuations of the equilibrium state, and it is clear that both initial states considered have relaxed to an equilibrium state with approximately the same average value for $\rm{var}(\tilde y)$. We also note that both systems end up with approximately the same condensate fraction of $f_{\rm cond}\approx0.15$, obtained from density matrix calculated by time-averaging states over the last 10 trap periods of the system evolution.
 
\section{Conclusions and outlook}
In this paper we have described an efficient spectral method for solving the projected Gross-Pitaevskii Equation for a Bose gas in a harmonic oscillator potential. 
We have shown that the nonlinear matrix elements can be calculated exactly using appropriately chosen quadrature grids and have described how this can be implemented as an efficient algorithm. We have also discussed numerous properties of our approach that allow various observables to be calculated, either in the spectral basis or by making appropriate use of quadrature grids. Finally, we have applied our method to an anisotropic 3D Bose gas to indicate typical features of the solutions.  

The rather unique requirement of the PGPE is that it  needs to be propagated on a small prescribed basis, usually containing of order a thousand modes. For this reason the additional computational cost of using a spectral method is offset by the rather small basis set size. However, recently our method has been used to simulate a trapped system with up to $4\times 10^5$  modes \cite{Wuster2007a}, and is reasonably efficient even for larger problems. However, there is much scope for development of the algorithm and we hope that the detailed discussion of numerics and example application in this paper will stimulate others in the community to develop more flexible and general PGPE algorithms

The algorithm outlined in this paper is an important step to realizing a comprehensive theory for simulating the dynamics of a trapped finite temperature Bose gas. The next steps will involve moving beyond the PGPE description to the Stochastic Gross-Pitaevskii Equation (SGPE) \cite{Gardiner2003a}, which includes the effects of coupling between the classical and incoherent regions. 
 This theory differs from the PGPE primarily through two new features: (i) Stochastic noise terms effect the evolution and cause the transfer of particles and energy into and out of the classical region. These contributions have already been added to the numerical approach outline here \cite{Davis2008SGPE}. (ii) Scattering terms corresponding to collisions between incoherent and classical region atoms that cause the transfer of momentum into the system. These terms are more challenging to implement and are the subject of current work. 
\section*{Acknowledgments} 
The author acknowledges a long and valuable collaboration with M.~J. Davis and many useful discussions with B.~I.  Schneider, N. Nygaard, A.~S. Bradley, A.~Bezett and T.~P. Simula. This work was financially supported by the Marsden Fund of New Zealand, the University of Otago and the New Zealand Foundation for Research Science and Technology under the contract NERF-UOOX0703: Quantum Technologies.

\end{document}